\documentclass{epl2} 

\newcommand{\citet}{\cite}
\newcommand{\citep}{\cite}

\makeatletter
\usepackage{pxfonts}

\makeatother

\begin{document}

\title{Thermodynamics of Extended Bodies in Special Relativity}

\author{Tadas K Nakamura}

\institute{                    
   Fukui Prefectural University, 910-1195 Fukui, JAPAN
}

\pacs{03.30.+p}{Special relativity}
\pacs{05.90.+m }{Other topics in statistical physics, thermodynamics,
and nonlinear dynamical systems}

\abstract{
 Relativistic thermodynamics is generalized to accommodate four dimensional
rotation in a flat spacetime. An extended body can be in equilibrium
when its each element moves along a Killing flow. There are three
types of basic Killing flows in a flat spacetime, each of which corresponds
to translational motion, spatial rotation, and constant linear acceleration;
spatial rotation and constant linear acceleration are regarded as
four dimensional rotation. Translational motion has been mainly investigated
in the past literature of relativistic thermodynamics. Thermodynamics
of the other two is derived in the present paper. 
}

\maketitle

\section{Introduction}

There have been published numerous papers on relativistic equilibrium
thermodynamics of extended bodies (bodies with finite volumes) in
a flat spacetime. Most of papers focused on a equilibrium of a body
with translational motion, i.e., each element of the body has the
same velocity. There has been a heated controversy on the relativistic
temperature with the translational motion (see, e.g.,\citet{yuen70AmJPh}
and references therein) in 1960s, and papers are still published to
this date (e.g.,\citet{landsberg-matsas04PhyA,aresdeparga05JPhA,tadas08arXiv,requardt08}).

The equilibrium temperature of an extended body in translational motion
is uniform within the body, in other words, local temperature at each
point is the same. In contrast, there can be equilibrium states in
which the local temperature is not uniform. Equilibrium with varying
local temperature was first examined by Tolman and Ehrenfest\citet{tolman-ehrenfest1930PhRv}
in the context of general relativity. With some assumptions they concluded
the local temperature should be inversely proportional to the square
root of the temporal component of the metric tensor. This result can
be derived from more general approach\citet{dixson78,Israel-Stewart1979AnPhy},
which shows equilibrium can take place only when the local inverse
temperature four vector satisfies the Killing equation.

This result in general relativity is of course applicable to flat
spacetimes. The local inverse temperature four vector depends on the
local velocities, and consequently the condition of equilibrium determines
the velocity distribution by the Killing equation. There are three
types of basic Killing vectors in a flat spacetime, each of which
corresponds to translational motion, spatial rotation, and constant
linear acceleration. Correspondingly there are three types of equilibrium
in a flat spacetime. The case of translational motion is trivial:
the uniform temperature. 

In the case of spatial rotation, the equilibrium state is determined
by the well known effect which is often expressed as {}``the rim
of a rotating wheel is hotter than the axis.'' This effect can be
interpreted as the result of centrifugal force acting on the energy.
The energy is equivalent to the mass in relativity and subject to
the centrifugal force, resulting higher energy density on the outer
side of a wheel. The same effect takes place in the case of constant
acceleration, resulting non-uniform local temperatures in the direction
of acceleration. \bigskip{}

The equilibrium with spacial rotation or constant acceleration is
a well known fact, however, past literature focused on its microscopic
aspect. In other words, the distribution of the local temperatures
has been mainly investigated, and little attention has been paid for
the global thermodynamical properties of extended bodies.

The basic strategy of thermodynamics is based on the fact that the
macroscopic properties of a body in equilibrium can be well represented
by a small number of thermodynamical parameters, such as temperature,
pressure, etc. It is trivial in non-relativistic thermodynamics that
the temperature can be expressed with one single value, since the
local temperatures are all equal in equilibrium. In contrast, there
can be equilibrium with non-uniform local temperatures as discussed
above, and it is not well known how to represent its global temperature
so far. 

The purpose of the present paper is to generalize the concept of global
temperature, or global inverse temperature more precisely, to be applicable
to the relativistic equilibrium with non-uniform local temperature.
Even when the local temperature varies, the condition of equilibrium
is a stringent constraint, which can determine thermodynamical state
uniquely by a small number of parameters, namely the generalized inverse
temperature.

The second law of thermodynamics will be extended to accommodate four
dimensional rotation (spatial rotation and constant acceleration)
with the generalized inverse temperature. This second law can tell
how the energy-momentum or angular momentum are transferred between
extended bodies thermodynamically. Further, by introducing the generalized
inverse temperature we can obtain a clear insight on the relativistic
thermal equilibrium. We can tell the similarity and difference of
the equilibrium states in translational motion, spatial rotation, and
constant acceleration.\bigskip{}

Non-relativistic thermodynamics is basically the theory of energetics.
The temperature is defined based on the conservation of the energy.
When we generalize it to relativity, energy must be treated as one
component of the energy momentum which is expressed by a contravariant
vector (1-vector). Consequently the inverse temperature becomes a
covariant four vector (1-form); this formulation was first proposed
by van Kampen \citet{kampen68PhRv} and later refined by Israel \citet{israel76AnPhy}.

The energy-momentum is a conserved quantity resulting from the translational
symmetry. There can be another conserved quantity, four dimensional
angular momentum namely, in a Minkowski spacetime resulting from the
rotational symmetry. As we will see in the present paper, there exists
another inverse temperature corresponding to the angular momentum.
In general, rotation can have six independent directions in a four
dimensional space. Consequently the four dimensional angular momentum
is a contravariant bivector (2-vector), with six components, and the
corresponding inverse temperature becomes a covariant bivector (2-form)
with six components. 

After deriving general expression for the inverse temperature, two
specific cases of four dimensional rotation will be examined in the
present paper. In a three dimensional space, any successive two rotations
can be combined into one rotation. This is not true for a four dimensional
space because there can be two independent rotations. For example,
a rotation in the $x_{0}x_{1}$ plane is independent of the rotation
in the $x_{2}x_{3}$ plane and the two cannot be combined into one
single rotation. In the present paper we treat the motion of a single
rotation; this is not general, but can clarify the essential thermodynamical
properties of four dimensional rotational motion with simplicity.
We examine two basic single rotations in a Minkowski spacetime: spatial
rotation and constant acceleration. With respect to the constant acceleration,
we further focus on the Rindler motion, which is one limited case
but important in applications.

\section{Temperature and Velocity in Equilibrium}

Let us suppose an extended body with a three volume $\Sigma$ is in
the equilibrium state, and examine its thermodynamical properties.
The local inverse temperature $\bar{\xi}(\bar{x})$ at each point
within the body is defined as\begin{equation}
\xi_{\mu}(\bar{x})=\frac{u_{\mu}(\bar{x})}{T(\bar{x})}\,,\label{eq:invtemp}\end{equation}
where $T(\bar{x})$ and $\bar{u}(\bar{x})$ are the local proper temperature
and the four velocity of the matter at a point $\bar{x}$ \citet{dixson78,Israel-Stewart1979AnPhy};
we denote a vector or a tensor as a whole by a bar, e.g., $\bar{\xi}$,
and its each component by a subscript or a superscript, e.g., $\xi_{\mu}$
in the present paper (precisely speaking, the position $\bar{x}$
may not be a vector, however, this expression does not cause confusion
in a flat spacetime). We use natural units $\hbar=c=G=k_{B}=1$ throughout
this paper unless otherwise stated.

When the body is in equilibrium, $\bar{\xi}$ satisfies the following
Killing equation \citet{dixson78,Israel-Stewart1979AnPhy}:\begin{equation}
\nabla_{\nu}\xi_{\mu}-\nabla_{\mu}\xi_{\nu}=0\,.\label{eq:killing}\end{equation}
The general solution to the above equation in a flat spacetime is
given as (see Appendix)

\begin{equation}
\xi_{\mu}=\beta_{\mu}+\lambda_{\mu\nu}(x^{\nu}-x_{0}^{\nu})\,,\label{eq:killing2}\end{equation}
where $\bar{\beta}$ is a constant vector, $\bar{x}_{0}$ is a certain
fixed point that corresponds to the center of four dimensional rotation,
and $\bar{\lambda}$ is a anti-symmetric tensor that satisfies $\lambda_{\mu\nu}=-\lambda_{\mu\nu}$
\citet{dixson78}. Both $\bar{\beta}$ and $\bar{\lambda}$ do not
depend on the position $\bar{x}$. The solution to (\ref{eq:killing})
has ambiguity in its amplitude since the Killing equation is linear;
the amplitude is determined so as to give the appropriate temperature
by (\ref{eq:invtemp}). When $\bar{\beta}$ , $\bar{\lambda}$ and
$\bar{x}_{0}$ are given, $T(\bar{x})$ and $\bar{u}(\bar{x})$ can
be calculated from (\ref{eq:invtemp}) as \begin{eqnarray}
T(\bar{x}) & = & |\beta_{\mu}+\lambda_{\mu\nu}(x^{\nu}-x_{0}^{\nu})|\,,\nonumber \\
u_{\mu}(\bar{x}) & = & T^{-1}(x)\,[\beta_{\mu}+\lambda_{\mu\nu}(x^{\nu}-x_{0}^{\nu})]\,.\label{eq:motion}\end{eqnarray}

Now that we obtain the equilibrium state, let us examine its thermodynamical
properties. Suppose there is an adiabatic energy-momentum supply to
the body, and the local energy momentum density increases by $\Delta T_{\rho}^{\mu}(\bar{x})$;
this is not uniform within the body in general. The local change of
the entropy four vector at a point $\bar{x}$ is given as $\Delta s_{\rho}=\xi_{\mu}\Delta T_{\rho}^{\mu}$
(see, e.g.,\citet{israel76AnPhy,maartens1996arXiv}), thus we can
write the total entropy change in the three volume $\Sigma$ as \begin{eqnarray}
\Delta S & = & \frac{1}{T_{0}}\int_{\Sigma}[\beta_{\mu}+\lambda_{\mu\nu}(x^{\nu}-x_{0}^{\nu})]\Delta T_{\rho}^{\mu}\, d\Sigma^{\rho}\nonumber \\
\, & = & \beta_{\mu}\Delta G^{\mu}+\lambda_{\mu\nu}\Delta M^{\mu\nu}\;.\label{eq:number5}\end{eqnarray}
In the above expression $\Delta\bar{G}$ and $\Delta\bar{M}$ are
the changes in energy-momentum and four dimensional angular momentum
defined by\begin{equation}
\Delta G^{\mu}=\int_{\Sigma}\Delta T_{\rho}^{\mu}\, d\Sigma^{\rho}\,,\;\;\;\Delta M^{\mu\nu}=\int_{\Sigma}(x^{\nu}-x_{0}^{\nu})\Delta T_{\rho}^{\mu}\, d\Sigma^{\rho}\,.\label{eq:momentum}\end{equation}
Precisely speaking, both $\Delta\bar{G}$ and $\Delta\bar{M}$ are
frame dependent when the body is not isolated\citet{tadas06PhLA}.
However, the dependence is canceled out when we calculate the entropy
and thus we do not pay attention to this point in the following.

Most of papers on relativistic thermodynamics of extended bodies assume
the case with $\bar{\lambda}=0$, i.e., for the translational motion
without rotation or acceleration. In this case van Kampen\citep{kampen68PhRv}
and Israel\citep{israel76AnPhy} suggested the concept of temperature
should be extended by defining a four vector $\beta_{\mu}$ as a inverse
temperature.

When $\bar{\lambda}\ne0$, we find the parameter $\bar{\lambda}$
in (\ref{eq:number5}) plays the same role to $\Delta\bar{M}$ as
$\bar{\beta}$ does to $\Delta\bar{G}$. Therefore, $\bar{\lambda}$
can be regarded as a thermodynamical parameter like the inverse temperature
four vector $\bar{\beta}$. Since $\bar{\lambda}$ is a bivector (2-form)
with six independent components, the inverse temperature has ten independent
components in total: four of $\beta_{\mu}$ and six of $\lambda_{\mu\nu}$.
This number corresponds to the number of independent Killing vector
fields in a Minkowski spacetime, or equivalently, the number of conserved
quantities resulting from the symmetry of spacetime.

The ten components of inverse temperature is the generalization of
the inverse temperature four vector; we can treat not only energy-momentum
but also four dimensional angular momentum with them. In the following
the combination of $(\bar{\beta},\bar{\lambda})$ is simply referred
as {}``inverse temperate''.

This inverse temperature has very basic function of the {}``temperature''
in thermodynamics, i.e., it determines how the irreversible exchange
of energy-momentum and angular momentum takes place spontaneously
as an thermodynamical effect. Suppose two isolated bodies each of
which is in equilibrium with inverse temperatures $\bar{(\beta}_{\textnormal{I}},\bar{\lambda}_{\textnormal{I}})$
and $\bar{(\beta}_{\textnormal{II}},\bar{\lambda}_{\textnormal{II}})$.
Then the relativistic second law may be stated in the following: the
thermal exchange of energy-momentum or angular momentum takes place
spontaneously only when the entropy increases, i.e., \begin{equation}
\Delta S=(\beta_{\textnormal{I}\mu}-\beta_{\textnormal{II}\mu})\Delta G^{\mu}+(\lambda_{\textnormal{I}\mu\nu}-\lambda_{\textnormal{II}\mu\nu})\Delta M^{\mu\nu}>0\,,\end{equation}
where $\Delta\bar{G}$ and $\Delta\bar{M}$ are the energy-momentum
and angular momentum transferred from the body II to body I. The two
bodies can be in total equilibrium only when all ten components of
the inverse temperatures are equal, which means the motions of the
two bodies belong to the same Killing vector field.

\section{Rotation and Acceleration}

The motion expressed in (\ref{eq:motion}) is a superposition of translational
motion and four dimensional rotation, and in general four dimensional
rotation can be divided into two categories: single rotation and double
rotation. In the following we concentrate on the single rotation for
simplicity. In the case of the Minkowski spacetime, the single rotation
can be further divided into two categories: spatial rotation and constant
acceleration. These two have similar mathematical structures, however,
their physical properties are considerably different.

The direction of a single rotation can be specified by a two dimensional
plane in which the rotation takes place. In a three dimensional space,
specifying a two dimensional plane is equivalent to specifying an
axis orthogonal to the plane. This does not work in a space with the
dimension higher than four because the orthogonal direction to a plane
is not unique, therefore, we need to specify the direction by a two
dimensional plane.

Two dimensional planes in a Minkowski spacetime are categorized into
two groups in general. One consists of the planes spanned by two spacelike
vectors, and planes in the other group are spanned by one timelike
and one spacelike vectors; the former defines the spatial rotation
and the latter defines the constant acceleration. We will examine
these two in the following.

\subsection{Spatial Rotation}

For spatial rotation, we can choose the $xy$ plane as the plane of
rotation without loss of generality, then we have $\lambda_{xy}\ne0$,
$\lambda_{ti}=\lambda_{yz}=\lambda_{zx}=0$ in (\ref{eq:killing2}).
We still have two degrees of freedom in the choice of axis directions
and can set $\beta_{y}=\beta_{z}=0$ with them. Then the killing vector
$\bar{\xi}$ in (\ref{eq:killing2}) can be written as \begin{equation}
\bar{\xi}=(\beta_{t},\beta_{x}+\lambda_{xy}(y-y_{0}),-\lambda_{xy}(x-x_{0}),0)\,.\label{eq:killing3}\end{equation}
The above expression can be simplified further by choosing the origin
as $x_{0}=0$ and $y_{0}=\beta_{x}/\lambda_{xy}$, which gives $\bar{\xi}=(\beta_{t},\lambda_{xy}y,-\lambda_{xy}x,0)$.
Then (\ref{eq:motion}) is reduced to \begin{equation}
\bar{u}=\frac{1}{\sqrt{1-\Omega^{2}r^{2}}}\,(1,\Omega y,-\Omega x,0,0)\,,\label{eq:rotation}\end{equation}
where $\Omega=\lambda_{xy}/\beta_{t}$ is the angular velocity and
$r=\sqrt{x^{2}+y^{2}}$ is the three dimensional distance from the
rotating axis. As well known, the motion must restricted within the
light cylinder, i.e., $r<\Omega^{-1}$ to keep the causality.

Then the inverse temperature is written as\begin{eqnarray}
\beta_{\mu} & = & \left\{ \begin{array}{cc}
T_{0}^{-1} & (\mu=t)\\
0 & (\textnormal{otherwise})\end{array}\right.\nonumber \\
\lambda_{\mu\nu} & = & \left\{ \begin{array}{cc}
\Omega T_{0}^{-1} & (\mu,\nu=x,y)\\
-\Omega T_{0}^{-1} & (\mu,\nu=y,x)\\
0 & \textnormal{(otherwise})\end{array}\right.\label{eq:invtemp1}\end{eqnarray}
where $T_{0}=T(\bar{x}_{0})$ is the temperature at the axis. The
local temperature becomes $T=T_{0}/\sqrt{1-\Omega r^{2}}$.

We understand from the above expression that the thermodynamical sate
of a rotating body is essentially determined by two parameters $\Omega$
and $T_{0}$, and the ten components of the inverse temperature are
derived from the Lorentz transform. The general expression of the
four velocity becomes\begin{equation}
\bar{u}=\frac{1}{\sqrt{1-\Omega^{2}r^{2}}}\,[U_{\mu}+\Omega_{\mu\nu}(x^{\nu}-x_{0}^{'\nu})]\,,\end{equation}
where $U_{\mu}=\beta_{\mu}/(\beta_{\rho}\beta^{\rho})$, $\Omega_{\mu\nu}=\lambda_{\mu\nu}/(\beta_{\rho}\beta^{\rho})$,
and $r$ is the three dimensional distance from the rotating axis.
The vector $\bar{U}$ represents the four velocity of the rotation
center, and it is perpendicular to the rotation plane, i.e., $U^{\mu}\Omega_{\mu\nu}q^{\nu}=0$
for any four vector $\bar{q}$. Note that the origin $x_{0}'$ is
not identical to $x_{0}$ in (\ref{eq:killing2}) because of the shift
$\beta_{x}/\lambda_{xy}$ to obtain (\ref{eq:rotation}).

\subsection{Constant Acceleration}

Constant acceleration is characterized by the rotating plane spanned
by one timelike and one spacelike vector, and we can choose the coordinate
such that the latter becomes the $tx$ plane, resulting $\lambda_{tx}\ne0$,
$\lambda_{ty}=\lambda_{tz}=\lambda_{ij}=0$. Further we can simplify
the expression with $\beta_{t}=\beta_{x}=0$ by choice of the origin
and $\beta_{z}=0$ by choice of the axis direction in the same way
as in the above subsection. 

Then the four velocity becomes\begin{equation}
\bar{u}=\frac{1}{\sqrt{\Omega^{2}\rho^{2}-1}}\,(\Omega x,\Omega t,1,0)\,,\label{eq:accel}\end{equation}
where $\rho^{2}=x^{2}-t^{2}$ and $\Omega=\lambda_{tx}/\beta_{y}$.
The motion must be in the region of $|\Omega\rho-1|>1$ to keep the
causality. The inverse temperature becomes\begin{eqnarray}
\beta_{\mu} & = & \left\{ \begin{array}{cc}
T_{0}^{-1} & (\mu=y)\\
0 & (\textnormal{otherwise})\end{array}\right.\nonumber \\
\lambda_{\mu\nu} & = & \left\{ \begin{array}{cc}
\Omega T_{0}^{-1} & (\mu,\nu=t,x)\\
-\Omega T_{0}^{-1} & (\mu,\nu=x,t)\\
0 & \textnormal{(otherwise})\end{array}\right.\label{eq:invtemp2}\end{eqnarray}

The above expressions are similar to (\ref{eq:rotation}): here the
spatial coordinate $y$ takes the place of the temporal coordinate
$t$ in (\ref{eq:rotation}). One difference to be noted is that a
case with $\beta_{y}=0$ ($T_{0},\Omega\rightarrow\infty$ with finite
$\Omega T_{0}^{-1}$) is allowed here. The four vector $\bar{u}$
must be time like, therefore, $\beta_{t}$ must be nonzero in the
spatial rotation. The four velocity of the acceleration, in contrast,
can be timelike even when $\beta_{y}=0$.

The motion of constant acceleration is usually investigated assuming
$\beta_{y}=0$ in the past literature. In this case (\ref{eq:accel})
becomes the well known Rindler motion:\begin{equation}
\bar{u}=\frac{1}{\rho}\,(x,t,0,0)\,,\end{equation}
It should be noted that the parameter $\Omega$ vanishes in the above
expression. This means the equilibrium velocity distribution of the
constant acceleration is uniquely determined in general without tuning
parameters. The acceleration at each trajectory is given as $a=1/\rho$,
in other words, the magnitude of the acceleration is determined by
the distance from the origin. 

In this case the local temperature is inversely proportional to $\rho$
i.e., $T(\rho)\propto1/\rho$, therefore it diverges at the origin.
This means the definition of the global inverse temperature based
on $T_{0}$, which was done in (\ref{eq:invtemp1}) or (\ref{eq:invtemp2}),
is inappropriate. The inverse temperature in this case may be written
using a new parameter $\Lambda=\Omega T_{0}^{-1}$as\begin{eqnarray}
\beta_{\mu} & = & 0\nonumber \\
\lambda_{\mu\nu} & = & \left\{ \begin{array}{cc}
\Lambda & (\mu,\nu=t,x)\\
-\Lambda & (\mu,\nu=x,t)\\
0 & \textnormal{(otherwise})\end{array}\right.\label{eq:rindtemp}\end{eqnarray}
Consequently the thermal property of the system is derived from only
one parameter $\Lambda$, in contrast to the two independent parameters
$\Omega$ and $T_{0}$ for the spatial rotation.

\section{Concluding Remarks}

Theory of relativistic thermodynamics is generalized to accommodate
four dimensional rotation in the present paper. The equilibrium state
of an extended body (body with finite volume) has been mainly investigated
in the past literature assuming the translational motion, i.e., the
velocity of each element vanishes in the comoving frame. There can
be other types of motion, spatial rotation and constant acceleration
namely, with which an equilibrium state is possible. These types of
motion can be regarded as four dimensional rotation, and equilibrium
is possible because of the rotational symmetry of the Minkowski spacetime.

From the non-relativistic theory of statistical mechanics we understand
the conventional temperature is the result of energy conservation
law. For relativistic translational motion, momentum also obeys its
conservation law and should be treated in the same way as energy.
This approach was adopted by van Kampen \citet{kampen68PhRv} and
Israel\citet{israel76AnPhy}, who proposed to treat the inverse temperature
as a four vector. The four components of the vector come from the
conservation laws of energy and three components of momentum. When
we generalize their theory to accommodate four dimensional rotation,
four dimensional angular momentum must be treated in the same way
as the energy-momentum because it is also a conserved quantity. 

Consequently there arises another kind of inverse temperature for
the angular momentum; this inverse temperature becomes covariant bivector
(2-form) since the angular momentum is contravariant bivector (2-vector).
Then the thermodynamical behavior of the body is completely determined
by two inverse temperatures with ten components in total. 

In the present paper the above formulation is applied to the motion
of single rotation, There are two types of single rotation in the
Minkowski spacetime; one is spatial rotation and the other is constant
acceleration. The result shows that the ten components are derived
from the two parameters, the global temperature $T_{0}$ and the angular
velocity $\Omega$ namely, by the Lorentz transform. 

In the case of spatial rotation, local temperature at each point of
the body is uniquely determined by these two parameters, in agreement
with the past literature: the rim of a rotating disk has higher temperature
than the center. Constant acceleration has the similar properties
in general. However, the case of Rindler motion, which is one special
case but especially important in applications, has a singular property.
The ten components of the inverse temperature is derived from only
one parameter $\Lambda=\Omega/T_{0}$, since we can eliminate the
inverse temperature of energy-momentum ($\bar{\beta}=0$). 

We make a brief comment on the relation of the present result to thermodynamics
of quantum vacua before closing this section. It is believed that
an observer with relativistic constant acceleration (Rindler motion)
finds a quantum vacuum thermalized with a certain temperature (see,
e.g., \citet{birrel-davies84}). This effect, which is called Unruh
effect, seem to contradict the result obtained in the present paper,
because the temperature of the Unruh effect cannot be derived from
our formulation. However, it should be noted that the Unruh effect
is a pure quantum effect; our result is within the framework of classical
thermodynamics which is valid when the quantum effect is negligible.

The local temperature of the Unruh effect is estimated as $T_{\textnormal{unruh}}=\hbar a/2\pi c$,
where $a$ is the acceleration of the Rindler observer \citet{birrel-davies84};
here we explicitly denote the speed of light by $c$. The effect is
negligible when $T_{\textnormal{unruh}}\ll T$ where $T$ the local
temperature in (\ref{eq:invtemp}). We can express this condition
in terms of the parameter $\Lambda$ in (\ref{eq:rindtemp}). The
global inverse temperature of the Unruh effect is derived from $\Lambda_{\textnormal{unruh}}=\hbar/2\pi$;
note that $\Lambda$ has the unit of $\hbar^{-1}$ because it is generalization
of inverse temperature that corresponds to four dimensional angular
momentum. Then the condition for the Unruh effect to be negligible
may be written as $\hbar\ll2\pi c/\Lambda$. This condition is satisfied
unless the temperature is extremely small because of the large factor
$c$, which means our result here can applicable for a wide variety
of relativistic phenomena hopefully.

%\revision{
{
\newcounter{section}\newcounter{subsection}\appendix

\section{Appendix}

It is easy to see the killing vector obtained in (\ref{eq:killing2})
satisfies the Killing equation (\ref{eq:killing}) by direct calculation,
however, its physical implication might not be clear to the readers
who are not familiar with the subject. We will give a brief explanation
in this Appendix.

When we choose a specific reference frame, the anti-symmetric tensor
$\lambda_{\mu\nu}$ in (\ref{eq:killing2}) can be decomposed into
the temporal and spatial parts as\begin{equation}
\bar{\lambda}=\bar{\lambda}^{(s)}+\bar{\lambda}^{(t)}=\left(\begin{array}{cccc}
0 & 0 & 0 & 0\\
0 & 0 & \lambda_{xy} & -\lambda_{xx}\\
0 & -\lambda_{xy} & 0 & \lambda_{yz}\\
0 & \lambda_{zx} & -\lambda_{yz} & 0\end{array}\right)+\left(\begin{array}{cccc}
0 & \lambda_{tx} & \lambda_{ty} & \lambda_{tz}\\
-\lambda_{tx} & 0 & 0 & 0\\
-\lambda_{ty} & 0 & 0 & 0\\
-\lambda_{tz} & 0 & 0 & 0\end{array}\right)\label{eq:a1}\end{equation}
The meaning of the first term can intuitively understood by expressing
the second term of (\ref{eq:killing2}) in the three dimensional form;\begin{equation}
\lambda_{\mu\nu}^{(s)}(x_{\mu}-x_{\mu}^{0})\rightarrow\boldsymbol{\lambda}\times(\mathbf{x}-\mathbf{x}_{0})\;\;\;\mu=x,y,z\end{equation}
where $\boldsymbol{\lambda}=(\lambda_{xy},\lambda_{yz},\lambda_{zx})$.
It is easy to see the three velocity calculated from the above expression
represents rotational motion when $\bar{\lambda}^{(t)}=0$. Note that
$\bar{\beta}$ must not vanish in this case because the four velocity
has to be timelike; the combination of $\bar{\lambda}$ and $\bar{\beta}$
determines the angular velocity of the rotation.\bigskip{}

The second term of (\ref{eq:a1}) has the same mathematical structure
as the first term, therefore, it should be understood as to correspond
to four dimensional rotation; the difference is in the point that the
plain of rotation is spanned by one timelike and one spacelike
vectors. It is known such rotation in a Minkowski spacetime represents
Lorentz transform, which means the change in velocity. Therefore, the
rotation at constant rate means the motion with constant acceleration.

To see the above point with actual calculation, let us examine a spacial 
case with $\lambda_{\mu\nu}^{(s)}=\beta_{\mu}=0$ as
an example. We can set $\lambda_{ty}=\lambda_{tz}=x_{\mu}^{0}=0$ with
an appropriate choice of the spatial axis. Suppose a {}``test
particle'' is moving along the killing flow in (\ref{eq:killing2})
whose four velocity is given by the second equation of
(\ref{eq:motion}).  Then its equation of motion may be written
as\begin{equation}
V_{x}(t)=\frac{dX(t)}{dt}=\frac{\lambda_{tx}t}{1+\lambda_{tx}X(t)}\end{equation}
The above equation gives the following velocity with the initial value
$V_{x}=0$ and $X=X_{0}$ at $t=0$\begin{equation}
  V_{x}(t)=\frac{\lambda_{tx}t}{\sqrt{(1+\lambda_{tx}X_{0})^{2}+\lambda_{tx}^{2}t^{2}}}\;,\end{equation}
When $\lambda_{tx}t,\lambda_{tx}X_{0}\ll1$, the three velocity $V_{x}$
is much smaller than the unity, which means the on-relativistic limit.
The above expression gives the non-relativistic constant acceleration
$v_{x}=\lambda_{tx}t$ in this limit. Thus we understand the motion
corresponds to $\bar{\lambda}^{(t)}$ is the relativistic
generalization of constant acceleration.

The motion defined by (\ref{eq:killing2}) {[}or (\ref{eq:motion})
equivalently] is the superposition of the translational motion, spatial
rotation, and constant acceleration in general. It should be noted
the decomposition in (\ref{eq:a1}) is not unique but dependent of
the choice of reference frame. We have examined the consequences of
$\bar{\lambda}^{(s)}$ and $\bar{\lambda}^{(s)}$ separately in this
appendix for simplicity, however, the general case may be complicated
and cannot be understood as a simple superposition of elements.

}

\bibliographystyle{iopart-num}
\bibliography{rel-thermo}

\end{document}